\documentclass{article}
\usepackage{amssymb, amsmath, amsfonts}
\usepackage{color}
\usepackage{amsmath}
\usepackage{amsfonts}
\usepackage{amssymb}
\usepackage{graphicx}%
\providecommand{\U}[1]{\protect\rule{.1in}{.1in}}
\numberwithin{equation}{section}
\newtheorem{theorem}{Theorem}
\newtheorem{lemma}[theorem]{Lemma}
\newtheorem{definition}[theorem]{Definition}

\begin{document}

\title{Bi-presymplectic representation of Liouville integrable systems and related
separability theory}
\author{Maciej B{\l }aszak}

\begin{titlepage}
\maketitle
\noindent
{\small Division of Mathematical Physics, A. Mickiewicz University, Umultowska 85 , 61-614 Poznan, Poland,}
{\small e-mail blaszakm@amu.edu.pl}\\
\begin{abstract}
Bi-presymplectic chains of one-forms of arbitrary co-rank are considered. The conditions in which such chains
represent some Liouville integrable systems and the conditions in which there exist related bi-Hamiltonian chains of vector fields are presented.
In order to derived the construction of bi-presymplectic chains, the notions of dual Poisson-presymplectic pair,
d-compatibility of presymplectic forms and d-compatibility of Poisson bivectors is used. The completely algorithmic construction
of separation coordinates is demonstrated. It is also proved that
St\"{a}ckel separable systems have bi-inverse-Hamiltonian representation, i.e. are represented by bi-presymplectic
chains of closed one-forms. The co-rank of related structures depends
on the explicit form of separation relations.
\end{abstract}
\end{titlepage}

\section{Introduction}

The theory of finite dimensional conservative integrable systems has a long
history, starting from the works of Lagrange, Hamilton and Jacobi in the first
half of XIX century. In fact the Hamilton-Jacobi (HJ) theory is one of the
most powerful methods of integration by quadratures a wide class of systems
described by nonlinear ordinary differential equations, with a long history as
a part of analytical mechanics. The theory in question is closely related to
the Liouville integrable Hamiltonian systems. The main difficulty of the HJ
approach is that it demands a distinguished coordinates, so called
\emph{separation coordinates}, in order to work effectively.

There are two efficient and systematic methods of construction of separation variables
for dynamical systems. The first one bases on Lax representation and r-matrix theory for
derivation of separation coordinates \cite{sk1}. In this approach the integrals of motion
in involution appear as coefficients of characteristic equation
(\emph{spectral curve}) of the Lax matrix. This method was successfully
applied for separating variables in many integrable systems \cite{sk1}-\cite{kuz}. The other one is a geometric separability theory on bi-Poissonian manifolds
 \cite{1}-\cite{m3}, related to the so-called
Gel'fand-Zakharevich (GZ) bi-Hamiltonian systems \cite{GZ,GZ1}. In this
approach the constants of motion are closely related to the so-called
\emph{separation curve} which is intimately related to the St\"{a}ckel
separation relations.

The bi-Poissonian formulation of finite dimensional integrable Hamiltonian
systems has been systematically developed for the last two decades (see
\cite{b1} and the literature quoted there). It has been found that most of the
known Liouville integrable finite dimensional systems have more then one
Hamiltonian representation. Moreover, in the majority of known cases, both
Poisson structures of a given flow are degenerated. For such systems, related
bi-Poissonian (bi-Hamiltonian) commuting vector fields belong to one or more
bi-Hamiltonian chains starting and terminating with Casimirs of respective
Poisson structures. The most important aspect of such a construction is its
relation to the geometric separability theory. Having a bi-Hamiltonian
representation of a given system, the sufficient condition for the existence
of separation coordinates is the reducibility of one of the Poisson structures
onto a symplectic leaf of the other one. Unfortunately, this procedure is
non-algoritmic and has to be considered independently from case to case.
Moreover, we do not have a proof that it is always possible for any GZ
system. Anyway, once the reduction is done, the remaining procedure of the
construction of separation coordinates is almost algoritmic. The relevance of
bi-Hamiltonian formalism in separability theory was recently confirmed in
\cite{bb1}, where it was proved that arbitrary St\"{a}ckel system, defined by
an appropriate separation relations, has a bi-Hamiltonian extension.

On the other hand, it is well known from the classical mechanics, that if the
Poisson structure is nondegenerate, i.e. if the rank of the Poisson tensor is
equal to the dimension of a phase space, then the phase space becomes a
symplectic manifold with a symplectic structure being just the inverse of the
Poisson structure. In such a case there exists an alternative (dual)
description of Hamiltonian vector fields in the language of symplectic
geometry. So, a natural question arises, whether one can construct such a dual
picture in the degenerated case, when there is no natural inverse of the
Poisson tensor \cite{d}. For such tensors the notion of dual presymplectic
structures was developed in \cite{bm,b}.

The presymplectic picture is especially interesting in the case of Liouville
integrable systems. As was mentioned above, there is well developed
bi-Hamiltonian theory of such systems, based on Poisson pencils of the GZ
type, with polynomial in pencil parameter Casimir functions and related
separability theory. The important question is whether it is possible to
formulate an independent, alternative bi-presymplectic (bi-inverse-Hamiltonian
in particular) theory of Liouville integrable systems with related
separability theory and how both theories are related to each other.

The advantage of formalism presented is as follows. In the bi-Hamiltonian approach the existence of bi-Hamiltonian representation of a given flow is a necessary condition of separability but not a sufficient one.
Contrary, the existence of bi-presymplectic representation of a flow considered is a sufficient condition of separability. Moreover, the construction of separation coordinates is a fully algorithmic procedure (in a generic case obviously), as the restriction of both presymplectic structures to any leaf of a given foliation always exists and is a simple task.
For this reason the new formalism presented in the paper seems to be relevant for the modern separability theory.

The present paper develops the general bi-presymplectic theory of Liouville
integrable systems when the co-rank of presymplectic forms is arbitrary. The
whole formalism is based on the notion of \emph{d-compatibility} of
presymplectic forms and \emph{d-compatibility} of Poisson bivectors. Some
elements of that formalism was presented in papers \cite{b,bgz}. Here we
present a complete picture. Finally it is
shown that any St\"{a}ckel system, defined by an appropriate separation
relations, has a bi-inverse-Hamiltonian representation, what confirms the
relevance of presented formalism.

The paper is organized as follows. In section 2 we give some basic information
on Poisson tensors, presymplectic two-forms, Hamiltonian and inverse
Hamiltonian vector fields and dual Poisson-presymplectic pairs. In sections 3
the concept of d-compatibility of Poisson bivectors and d-compatibility of
closed two-forms is developed. Then, in section 4, the main properties of
bi-presymplectic chains of arbitrary co-rank are investigated. The conditions
in which the bi-presymplectic chain is related to some Liouville integrable
system and the conditions in which the chain is bi-inverse-Hamiltonian are
presented. Moreover, the conditions in which Hamiltonian vector fields,
constructed from a given bi-presymplectic chain, constitute a related
bi-Hamiltonian chain are also found. In section 5 we prove that arbitrary
St\"{a}ckel system, defined by an appropriate set of separation relations, has
a bi-inverse-Hamiltonian formulation. Finally, in section 6, we illustrate
presented theory by few representative examples.

Our treatment in this work is local.  Thus,
we always restrict our considerations  to the domain $\mathcal{O}$ of manifold $M$ where appropriate functions, vector
fields and one-forms never vanish and respective Poisson tensors and presymplectic forms are of constant co-rank.
In some examples we perform calculations in particular local chart from $\mathcal{O}$.

\section{Preliminaries}

Given a manifold $\mathcal{M}$ of $\dim\mathcal{M}=m,$ a \emph{Poisson
operator} $\Pi$ of co-rank $r$ on $\mathcal{M}$ is a bivector $\Pi\in
\Lambda^{2}(\mathcal{M})$ with vanishing Schouten bracket:
\begin{equation}
\lbrack\Pi,\Pi]_{S}=0, \label{Schouten}%
\end{equation}
whose kernel is spanned by exact one-forms
\[
\ker\Pi=Sp\{dc_{i}\}_{i=1,...,r}.
\]
In a local coordinate system $(x^{1},\ldots,x^{m})$ on $\mathcal{M}$ we have
\[
\Pi=\sum\limits_{i<j}^{m}\Pi^{ij}\frac{\partial}{\partial x^{i}}\wedge
\frac{\partial}{\partial x^{j}},
\]
while the Poisson property (\ref{Schouten}) takes the form
\[
\sum_{l}(\Pi^{lj}\partial_{l}\Pi^{ik}+\Pi^{il}\partial_{l}\Pi^{kj}+\Pi
^{kl}\partial_{l}\Pi^{ji})=0,\ \ \ \partial_{i}:=\frac{\partial}{\partial
x^{i}}.
\]
Let $C(\mathcal{M})$ denote the space of all smooth real-valued functions on
$\mathcal{M}$. A function $c\in C(\mathcal{M})$ is called the \emph{Casimir
function} of the Poisson operator $\Pi$ if $\Pi dc=0$. Having a Poisson tensor
we can define a Hamiltonian vector fields on $\mathcal{M}$. A vector field
$X_{F}$ related to a function $F\in C(\mathcal{M})$ by the relation
\begin{equation}
X_{F}=\Pi dF,
\end{equation}
is called the \emph{Hamiltonian vector field} with respect to the Poisson
operator $\Pi$.

A linear combination $\Pi_{\lambda}=\Pi_{1}+\lambda\Pi_{0}$ ($\lambda
\in\mathbb{R}$) of two Poisson operators $\Pi_{0}$ and $\Pi_{1}$ is called a
\emph{Poisson pencil} if the operator $\Pi_{\lambda}$ is Poisson for any value
of the parameter $\lambda$. In this case we say that $\Pi_{0} $ and $\Pi_{1}$
are \emph{compatible}$.$ When all Casimir functions of $\Pi_{\lambda}$ are
polynomials in parameter $\lambda$ then we say that the pencil is of
Gel'fand-Zakharevich (GZ) type.

Further, a \emph{presymplectic operator} $\Omega$ on $\mathcal{M}$ is defined
by a two-form that is closed, i.e. $d\Omega=0,$ degenerated in general. In the
local coordinate system $(x^{1},\ldots,x^{m})$ on $\mathcal{M}$ we can
represent $\Omega$ as
\[
\Omega=\sum\limits_{i<j}^{m}\Omega_{ij}dx^{i}\wedge dx^{j},
\]
where the closeness condition takes the form
\[
\partial_{i}\Omega_{jk}+\partial_{k}\Omega_{ij}+\partial_{j}\Omega_{ki}=0.
\]
Moreover, the kernel of any presymplectic form is an integrable distribution.
\ A vector field $X^{F}$ related to a function $F\in C(\mathcal{M})$ by the
relation%
\begin{equation}
\Omega X^{F}=dF \label{1.2}%
\end{equation}
is called the \emph{inverse Hamiltonian vector field} with respect to the
presymplectic operator $\Omega$.

As in the case of presymplectic forms their linear combination is always
presymplectic, hence the notion of compatibility, as it was defined for
Poisson tensors, does not make sense. We will come back to this problem in the
next section.

Any non-degenerate closed two form on $\mathcal{M}$ is called a
\emph{symplectic} form. The inverse of a symplectic form is an
\emph{implectic} operator, i.e. invertible Poisson tensor on $\mathcal{M}$ and
vice versa.

\begin{definition}
A pair $(\Pi,\,\Omega)$ is called dual implectic-symplectic pair on
$\mathcal{M}$ if $\Pi$ is non-degenerate Poisson tensor, $\Omega$ is
non-degenerate closed two-form and the following partition of unity holds on
$T\mathcal{M}$, respectively on $T^{\ast}\mathcal{M}$: $I=\Pi\Omega$ and
$I=\Omega\Pi.$
\end{definition}

So, in the non-degenerate case, dual implectic-symplectic pair is a pair of
mutually inverse operators on $\mathcal{M}$. Moreover, the Hamiltonian and the
inverse Hamiltonian representations are equivalent as for any implectic
bivector $\Pi$ there is a unique dual symplectic form $\Omega=\Pi^{-1}$ and
hence a vector field Hamiltonian with respect to $\Pi$ is an inverse
Hamiltonian with respect to $\Omega$.

Let us extend these considerations onto a degenerate case. In order to do it
let us introduce the concept of dual pair as it was done in \cite{bm}.
Consider a manifold $\mathcal{M}$ of an arbitrary dimension $m$.

\begin{definition}
\cite{bm} A pair of tensor fields $(\Pi,\Omega)$ on $\mathcal{M}$ of co-rank
$r$, where $\Pi$ is a Poisson tensor and $\Omega$ is a closed two-form, is
called a dual pair (Poisson-presymplectic pair) if there exists $r$ one-forms
$dc_{i}$ and $r$ linearly independent vector fields $Z_{i}$, such that the
following conditions are satisfied:\newline1. $\ker\Pi=Sp\{dc_{i}:\,i=1,\dots r\}$.\newline2.
$\ker\Omega=Sp\{Z_{i}:\,i=1,\dots r\}$.\newline3. $Z_{i}(c_j)=\delta_{ij}$,
$i=1,2\dots r$.\newline4. The following partition of
unity holds on $T\mathcal{M}$, respectively on $T^{\ast}\mathcal{M}$
\begin{equation}
I=\Pi\Omega+\sum_{i=1}^{r}Z_{i}\otimes dc_{i},\qquad I=\Omega\Pi+\sum
_{i=1}^{r}dc_{i}\otimes Z_{i}, \label{3}%
\end{equation}
where $\otimes$ denotes the tensor product.
\end{definition}

A presymplectic form $\Omega$ plays the role of an 'inverse' of Poisson
bivector $\Pi$ in the sense that on any symplectic leaf of the foliation
defined by $\ker\Pi$, the restrictions of $\Omega$ and $\Pi$ are inverses of
each other. More information on geometric interpretation of dual pairs the
reader can find in \cite{bm}. Contrary to the non-degenerated case, for a
given Poisson tensor $\Pi$ the choice of its dual is not unique. Also for a
given presymplectic form $\Omega$ the choice of dual Poisson tensor is not
unique. We will come back to that problem at the end of this section.

For the degenerate case the Hamiltonian and the inverse Hamiltonian vector
fields are defined in the same way as for the non-degenerate case, but for
degenerate structures the notion of Hamiltonian and inverse Hamiltonian vector
fields do not coincide. For any degenerate dual pair it is possible to find a
Hamiltonian vector field that is not inverse Hamiltonian and an inverse
Hamiltonian vector field that is not Hamiltonian. Actually, assume that
$(\Pi,\Omega)$ is a dual pair, $X_{F}=\Pi dF$ is a Hamiltonian vector field
and $dF=\Omega X^{F}$ is an inverse Hamiltonian one-form, where $X^{F}$ is an
inverse Hamiltonian vector field. Having applied $\Omega$ to both sides of
Hamiltonian vector field, $\Pi$ to both sides of inverse Hamiltonian one-form
and using the decomposition (\ref{3}) we get
\begin{equation}
dF=\Omega(X_{F})+\sum_{i=1}^{r}Z_{i}(F)dc_{i},\qquad X_{F}=X^{F}%
-\sum_{i=1}^{r}X^{F}(c_i)Z_{i}.
\end{equation}
It means that an inverse Hamiltonian vector field $X^{F}$ is simultaneously a
Hamiltonian vector field $X_{F}$, i.e. $X^{F}=X_{F}$, if $dF$ is annihilated
by $\ker(\Omega)$ and $X^{F}$ is annihilated by $\ker(\Pi)$. Moreover, for any
dual pair $(\Pi,\Omega)$, the following important relations hold \cite{bm}
\begin{equation}
\lbrack Z_{i},Z_{j}]=0,\quad L_{X_{F}}\Pi=0,\quad L_{Z_{i}}\Pi=0,\quad
L_{X^{F}}\Omega=0,\quad L_{Z_{i}}\Omega=0,
\end{equation}
for $i,j=1,...,r,$ where $L_{X}$ is the Lie-derivative operator in the
direction of vector field $X$ and $[.\,,.]$ is a commutator.

Let us return to a 'gauge freedom' for a duality property. In other words:
given a dual pair $(\Pi,\Omega)$ how can we deform $\Omega$ to a new
presymplectic form $\Omega^{\prime}$ so that $(\Pi,\Omega^{\prime})$ is again
a dual pair, or how can we deform $\Pi$ to a new Poisson operator $\Pi
^{\prime}$ so that $(\Pi^{\prime},\Omega)$ is also a dual pair?

\begin{lemma}
\cite{bm} Let $\Pi$ be a fixed Poisson tensor and $\Omega$ be a dual
presympectic form. Assume that $dc_{i}\in\ker\Pi$, $Z_{i}\in
\ker\Omega$ and $Z_{i}(c_{j})=\delta_{ij}$. Define
\[
\Omega^{\prime}=\Omega+%
{\textstyle\sum\limits_{i}}
df_{i}\wedge dc_{i},
\]
where $f_{i}$ $\in C(M)$. Then $(\Pi,\Omega^{\prime})$ is a dual pair, with
$\ker(\Omega^{\prime})=$ $Sp\left\{  Z_{i}^{\prime}=Z_{i}-\Pi\,df_{i}\right\}
$, provided that
\begin{equation}
Z_{i}(f_{j})-Z_{j}(f_{i})+\Pi(df_{i},df_{j})=0\ \text{ for all \ }i,j\text{. }
\label{gaugewar}%
\end{equation}

\end{lemma}

\begin{lemma}
\cite{bm} Let $\Omega$ be a fixed presymplectic form and $\Pi$ be a dual
Poisson tensor. Assume that $Z_{i}\in\ker\Omega$, $dc_{i}\in\ker
\Pi$ and $Z_{i}(c_{j})=\delta_{ij}$. Define
\begin{equation}
\Pi^{\prime}=\Pi+%
{\textstyle\sum\limits_{i}}
Z_{i}\wedge K_{i},
\end{equation}
where $K_{i}$ are vector fields such that
\begin{equation}
K_{i}=\Pi dF_{i},\quad dF_{i}=\Omega K_{i}\quad\Rightarrow\quad Z_{j}%
(F_{i})=0,\quad K_{j}(c_{i})=0,\ \ \ i,j=1,...,r,
\end{equation}
for some functions $F_{i}$ $\in C(M)$. Then, $(\Pi^{\prime},\Omega)$ is a dual
pair, with $\ker(\Pi^{\prime})=Sp\left\{  dc_{i}^{\prime}\right\}
,\ c_{i}^{\prime}=c_{i}+F_{i},$ provided that
\[
\Omega(K_{i},K_{j})=0\ \ \text{ for all }i,j.
\]

\end{lemma}

Poisson tensor $\Pi,$ considered as the mapping $\Pi:T^{\ast}\mathcal{M}%
\rightarrow T\mathcal{M}$, induces a Lie bracket on the space $C(\mathcal{M}%
)$
\begin{equation}
\left\{  .,.\right\}  _{\Pi}:C(\mathcal{M})\times C(\mathcal{M})\rightarrow
C(\mathcal{M})\text{, \ }\left\{  F,G\right\}  _{\Pi}\overset{\mathrm{def}}%
{=}\left\langle dF,\Pi\,dG\right\rangle =\Pi(dF,dG), \label{bracket}%
\end{equation}
(where $\left\langle .,.\right\rangle $ is the dual map between $T\mathcal{M}$
and $T^{\ast}\mathcal{M}$) which is skew-symmetric and satisfies Jacobi
identity. It is called a \emph{Poisson bracket}. Jacobi identity for
(\ref{bracket}) follows from the property (\ref{Schouten}) of $\Pi.$

When a Poisson operator $\Pi$ is nondegenerate its dual $\Omega$ is its
inverse $\Omega=\Pi^{-1}.$ Moreover, any Hamiltonian vector field with respect
to $\Pi$ is simultaneously the inverse Hamiltonian with respect to $\Omega$
and $X_{F}=X^{F}.$ Hence, a symplectic operator $\Omega$ defines the same
Poisson bracket as the related Poisson operator $\Pi$
\begin{align}
\{F,G\}^{\Omega}  &  \overset{\mathrm{def}}{=}\Omega(X_{F},X_{G})=<\Omega
X^{F},X_{G}>=<dF,X_{G}>=<dF,\Pi dG>\label{brackett}\\
&  =\Pi(dF,dG)=\{F,G\}_{\Pi}.\nonumber
\end{align}

What is important, when $\Pi$ is a degenerate Poisson tensor and $\Omega$ is
its an arbitrary dual two-form, the formula (\ref{brackett}) is still valid.
It follows from the fact that although $X_{F}\neq X^{F}$, but $<\Omega
X_{F},\Pi dG>=<\Omega X^{F},\Pi dG>$.

Finally, we remind the reader two identities important for further
considerations. Let $\Pi$ be a Poisson bivector and $\Omega$ be a closed
two-form, then
\begin{equation}
L_{\Pi\gamma}\Pi+\Pi d\gamma\Pi=0,\ \ \ \ \ \ L_{\tau}\Omega=d(\Omega\tau),
\label{2.5}%
\end{equation}
where $\tau\in T\mathcal{M}$ and $\gamma\in T^{\ast}\mathcal{M}$.

\section{D-compatibility of closed two-forms and Poisson bivectors}

In the following section we develope a concept of d-compatibility which is
crutial for our further considerations. Let us start with a non degenerate case.

\begin{definition}
We say that a closed two-form $\Omega_{1}$ is d-compatible with a symplectic
form $\Omega_{0}$ if $\Pi_{0}\Omega_{1}\Pi_{0}$ is a Poisson tensor and
$\Pi_{0}=\Omega_{0}^{-1}$ is dual to $\Omega_{0}$.
\end{definition}

\begin{definition}
We say that a Poisson tensor $\Pi_{1}$ is d-compatible with an implectic
tensor $\Pi_{0}$ if $\Omega_{0}\Pi_{1}\Omega_{0}$ is closed and $\Omega
_{0}=\Pi_{0}^{-1}$ is dual to $\Pi_{0}$.
\end{definition}

Now, the following lemma relates d-compatible Poisson structures, of which one
is implectic, and d-compatible closed two-forms, of which one is symplectic.

\begin{lemma}
\cite{bgz}

\item Let $(\Pi_{0},\Omega_{0})$ be a dual implectic-symplectic pair.\vspace
{0.4cm}\newline{(i)} Let a Poisson tensor $\Pi_{1}$ be d-compatible with
$\Pi_{0}$. Then $\Omega_{0}$ and $\Omega_{1}=\Omega_{0}\Pi_{1}\Omega_{0}$ are
d-compatible closed two-forms.\vspace{0.4cm}\newline{(ii)} Let a closed
two-form $\Omega_{1}$ be d-compatible with $\Omega_{0}$. Then $\Pi_{0}$ and
$\Pi_{1}=\Pi_{0}\Omega_{1}\Pi_{0}$ are d-compatible Poisson tensors.
\end{lemma}

Let us extend the notion of d-compatibility onto the degenerate case.

\begin{definition}
A closed two-form $\Omega_{1}$ is d-compatible with a closed two-form
$\Omega_{0}$ if there exists a Poisson tensor $\Pi_{0}$, dual to $\Omega_{0}$,
such that $\Pi_{0}\Omega_{1}\Pi_{0}$ is Poisson. Then we say that the pair
$(\Omega_{0},$ $\Omega_{1})$ is d-compatible with respect to $\Pi_{0}$.
\end{definition}

\begin{definition}
A Poisson tensor $\Pi_{1}$ is d-compatible with a Poisson tensor $\Pi_{0}$ if
there exists a presymplectic form $\Omega_{0}$, dual to $\Pi_{0}$, such that
$\Omega_{0}\Pi_{1}\Omega_{0}$ is closed. Then we say that the pair $(\Pi_{0},$
$\Pi_{1})$ is d-compatible with respect to $\Omega_{0}$.
\end{definition}

Comparing the notions of compatibility and d-compatibility for Poisson pair
$(\Pi_{0},\Pi_{1})$ we will show that when $\Pi_{0}$ is non-degenerated both
notions are equivalent, but for a degenerate case the notion of
d-compatibility is the stronger one. Actually, let us consider the following
identity, proved in \cite{b},
\begin{align}
&  L_{(\Pi_{1}+\lambda\Pi_{0})\gamma}(\Pi_{1}+\lambda\Pi_{0})+(\Pi_{1}%
+\lambda\Pi_{0})d\gamma(\Pi_{1}+\lambda\Pi_{0})\nonumber\\
&  =\lambda\{L_{\tau}(\Omega_{0}\Pi_{1}\Omega_{0})-d(\Omega_{0}\Pi_{1}%
\Omega_{0}\tau)-%
{\textstyle\sum\nolimits_{i}}
[\Omega_{0}(L_{Z_{i}}\Pi_{1})\Omega_{0}]\tau\wedge dc_{i}\label{bbb}\\
&  \ \ \ -%
{\textstyle\sum\nolimits_{i}}
\tau(c_{i})\Omega_{0}(L_{Z_{i}}\Pi_{1})\Omega_{0}\},\nonumber
\end{align}
where $\Pi_{0},\Pi_{1}$ are Poisson tensors, $(\Pi_{0},\Omega_{0})$ is a dual
pair, where $dc_{i}\in\ker\Pi_{0}$, $Z_{i}\in\ker\Omega_{0}$, $\tau\in
T\mathcal{M}$ and $\gamma=\Omega_{0}\tau\in T^{\ast}\mathcal{M}$. Assume first
that $\Pi_{0}$ and $\Pi_{1}$ are d-compatible with respect to $\Omega_{0}.$
Then $\Omega_{0}\Pi_{1}\Omega_{0}$ is closed and
\begin{equation}
L_{\tau}(\Omega_{0}\Pi_{1}\Omega_{0})-d(\Omega_{0}\Pi_{1}\Omega_{0}%
\tau)=0,\ \ \ \ \ \ \ \ \tau\in T\mathcal{M}. \label{3.3}%
\end{equation}
In particular, for $\tau=Z_{i}$, relation (\ref{3.3}) gives
\begin{equation}
\Omega_{0}(L_{Z_{i}}\Pi_{1})\Omega_{0}=0,\ \ \ \ \ \ \ \ \ \ i=1,...,r.
\label{3.4}%
\end{equation}
Hence
\begin{equation}
L_{(\Pi_{1}+\lambda\Pi_{0})\gamma}(\Pi_{1}+\lambda\Pi_{0})+(\Pi_{1}+\lambda
\Pi_{0})d\gamma(\Pi_{1}+\lambda\Pi_{0})=0 \label{3.5}%
\end{equation}
and $\Pi_{1}+\lambda\Pi_{0}$ is Poisson. On the other hand, from the
compatibility relation (\ref{3.5}) the d-compatibility (\ref{3.3}) follows
under additional conditions (\ref{3.4}).

\begin{theorem}
\label{theorem1}Let a Poisson tensor $\Pi_{0}$ and a closed two-form
$\Omega_{0}$ form a dual pair, where $Y_{0}^{(k)}\in\ker\Omega_{0}$,
$\,dH_{0}^{(k)}\in\ker\Pi_{0}$ and $Y_{0}^{(k)}(H_{0}^{(m)})=\delta
_{km},\ k,m=1,...,r$.\vspace{0.4cm}\newline(i) If $\Pi_{1}$ is a Poisson
tensor d-compatible with $\Pi_{0}$ with respect to $\Omega_{0}$, then forms
$\Omega_{0}$ and $\Omega_{1}=\Omega_{0}\Pi_{1}\Omega_{0}$ are
d-compatible with respect to $\Pi_0$.\vspace{0.4cm}\newline(ii) If $\Omega_{1}$ is a closed two-form
d-compatible with $\Omega_{0}$ with respect to $\Pi_{0}$, then Poisson tensors
$\Pi_{0}$ and $\Pi_{1}=\Pi_{0}\Omega_{1}\Pi_{0}$ are d-compatible with respect to $\Omega_{0}$, provided
that
\begin{equation}
\Pi_{0}\Omega_{1}Y_{0}^{(k)}=\Pi_{0}dF^{(k)},\ \ \ \ k=1,...,r \label{3.6}%
\end{equation}
for some functions $F^{(k)}\in C(\mathcal{M})$ and
\begin{equation}
\Omega_{1}(Y_{0}^{(k)},Y_{0}^{(m)})+Y_{0}^{(k)}(F^{(m)})-Y_{0}^{(m)}%
(F^{(k)})=const,\ \ \ \ \ \ \ \ k,m=1,...,r. \label{3.7}%
\end{equation}

\end{theorem}

\textbf{Proof.}\newline(i) $\Omega_{1}$ is closed as $\Pi_{1}$ is d-compatible
with $\Pi_{0}$. Then, $\Pi_{0}\Omega_{1}\Pi_{0}=\Pi_{0}\Omega_{0}\Pi_{1}%
\Omega_{0}\Pi_{0}$ is Poisson (as was shown in \cite{b}).\vspace
{0.3cm}\newline(ii) From the d-compatibility of $\Omega_{0}$ and $\Omega_{1}$
it follows that $\Pi_{1}$ is Poisson. Then,
\begin{align*}
\Omega_{0}\Pi_{1}\Omega_{0}  &  =\Omega_{0}\Pi_{0}\Omega_{1}\Pi_{0}\Omega
_{0}=(I-%
{\textstyle\sum\nolimits_{k}}
dH_{0}^{(k)}\otimes Y_{0}^{(k)})\Omega_{1}(I-%
{\textstyle\sum\nolimits_{m}}
\,Y_{0}^{(m)}\otimes dH_{0}^{(m)})\\
&  =\Omega_{1}+%
{\textstyle\sum\nolimits_{k}}
\,dH_{0}^{(k)}\wedge\Omega_{1}(Y_{0}^{(k)})+\frac{1}{2}%
{\textstyle\sum\nolimits_{k,m}}
\Omega_{1}(Y_{0}^{(m)},Y_{0}^{(k)})dH_{0}^{(k)}\wedge dH_{0}^{(m)}.
\end{align*}
From the assumption (\ref{3.6}) and decompositions (\ref{3}) it follows that
\[
\Omega_{1}Y_{0}^{(k)}=dF^{(k)}+%
{\textstyle\sum\nolimits_{m}}
\left[  \Omega_{1}(Y_{0}^{(k)},Y_{0}^{(m)})-Y_{0}^{(m)}(F^{(k)})\right]
dH_{0}^{(m)},
\]
hence,
\begin{align*}
\Omega_{0}\Pi_{1}\Omega_{0}  &  =\Omega_{1}+%
{\textstyle\sum\nolimits_{k}}
dH_{0}^{(k)}\wedge dF^{(k)}\\
&  \ \ \ +%
{\textstyle\sum\nolimits_{k,m}}
\left[  \frac{1}{2}\Omega_{1}(Y_{0}^{(k)},Y_{0}^{(m)})-Y_{0}^{(m)}%
(F^{(k)})\right]  dH_{0}^{(k)}\wedge dH_{0}^{(m)}\\
&  =\Omega_{1}+%
{\textstyle\sum\nolimits_{k}}
dH_{0}^{(k)}\wedge dF^{(k)}\\
&  \ \ \ +\frac{1}{2}%
{\textstyle\sum\nolimits_{k,m}}
\left[  \Omega_{1}(Y_{0}^{(k)},Y_{0}^{(m)})-Y_{0}^{(m)}(F^{(k)})+Y_{0}%
^{(k)}(F^{(m)})\right]  dH_{0}^{(k)}\wedge dH_{0}^{(m)}%
\end{align*}
and under condition (\ref{3.7}) $\Omega_{0}\Pi_{1}\Omega_{0}$ is closed.
$\Box$

The important for further considerations special case occurs when
\begin{equation}
\Omega_{1}(Y_{0}^{(k)},Y_{0}^{(m)})=0,\ \ \ \ Y_{0}^{(k)}(F^{(m)})=Y_{0}%
^{(m)}(F^{(k)}),\ \ \ \ k,m=1,...,r.\ \label{3.8}%
\end{equation}

\begin{theorem}
\label{theorem2} Let a Poisson tensor $\Pi_{0}$ and a closed two-form
$\Omega_{0}$ form a dual pair, where $Y_{0}^{(k)}\in\ker\Omega_{0}$,
$\,dH_{0}^{(k)}\in\ker\Pi_{0}$ and $Y_{0}^{(k)}(H_{0}^{(m)})=\delta
_{km},\ k,m=1,...,r$.\vspace{0.4cm}\newline(i)If $\Pi_{1}$ is a Poisson tensor
d-compatible with $\Pi_{0}$ with respect to $\Omega_{0}$ and
\begin{equation}
X^{(k)}=\Pi_{1}dH_{0}^{(k)}=\Pi_{0}dH_{1}^{(k)},\ \ \ \ \ \ \ k=1,...,r
\label{3.9}%
\end{equation}
are bi-Hamiltonian vector fields for some functions $H_{1}^{(k)}$, then
$\Omega_{0}$ and $\Omega_{1}=\Omega_{0}\Pi_{1}\Omega_{0}+%
{\textstyle\sum\nolimits_{k}}
dH_{1}^{(k)}\wedge dH_{0}^{(k)}$ is d-compatible pair of presymplectic forms
with respect to $\Pi_{0}$.\vspace{0.3cm}\newline(ii) If $\Omega_{1}$ is a
presymplectic form d-compatible with $\Omega_{0}$ with respect to $\Pi_{0}$
and
\begin{equation}
\beta^{(k)}=\Omega_{0}Y_{1}^{(k)}=\,\Omega_{1}Y_{0}^{(k)}%
,\ \ \ \ \ \ k=1,...,r \label{3.10}%
\end{equation}
are bi-presymplectic one-forms, then Poisson tensors $\Pi_{0}$ and $\Pi_{1}=\Pi_{0}\Omega
_{1}\Pi_{0}+%
{\textstyle\sum\nolimits_{k}}
X^{(k)}\wedge Y_{0}^{(k)}$, are d-compatible with respect to $\Omega_{0}$ if
\begin{equation}
X^{(k)}=\Pi_{0}\Omega_{1}Y_{0}^{(k)}=\Pi_{0}dF^{(k)},\quad\Pi_{0}\Omega
_{1}Y_{1}^{(k)}=\Pi_{0}dG^{(k)},\ \ \ \ \ \ k=1,...,r, \label{3.11}%
\end{equation}
for some functions $F^{(k)},G^{(k)}\in C(\mathcal{M})$ and
\begin{equation}
\Omega_{0}(Y_{1}^{(k)},Y_{1}^{(m)})=0,\ \ \ \ \ \ k,m=1,...,r, \label{3.12}%
\end{equation}%
\begin{equation}
Y_{0}^{(m)}(F^{(k)})=Y_{0}^{(k)}(F^{(m)}),\ \ \ \ \ \ k,m=1,...,r,
\label{3.13}%
\end{equation}%
\begin{equation}
Y_{1}^{(m)}(H_{0}^{(k)})=Y_{0}^{(k)}(F^{(m)}),\ \ \ \ \ \ k,m=1,...,r.
\label{3.14}%
\end{equation}

\end{theorem}

\textbf{Proof.}\newline(i) $\Omega_{1}$ is closed as $\Pi_{1}$ is d-compatible
with $\Pi_{0}$. Then, $\Pi_{0}\Omega_{1}\Pi_{0}=\Pi_{0}\Omega_{0}\Pi_{1}%
\Omega_{0}\Pi_{0}$ is Poisson (as was shown in \cite{b}).\vspace
{0.3cm}\newline(ii) From (\ref{3.11}) we have
\begin{equation}
\Omega_{1}(Y_{0}^{(k)},Y_{0}^{(m)})=0 \label{3.16}%
\end{equation}
and by previous theorem part (ii) the form $\Omega_{0}\Pi_{1}\Omega_{0}%
=\Omega_{0}\Pi_{0}\Omega_{1}\Pi_{0}\Omega_{0}$ is closed under condition
(\ref{3.13}). Moreover, (\ref{3.16}) yields
\[
\Omega_{0}Y_{1}^{(k)}=dF^{(k)}-%
{\textstyle\sum\nolimits_{m}}
Y_{0}^{(k)}(F^{(k)})dH_{0}^{(m)},
\]%
\[
\Omega_{1}Y_{1}^{(k)}=dG^{(k)}-%
{\textstyle\sum\nolimits_{m}}
Y_{0}^{(k)}(G^{(k)})dH_{0}^{(m)}.
\]
Conditions (\ref{3.12}) and (\ref{3.14}) are sufficient for $\Pi_{1}$ to be a
Poisson tensor. From (\ref{3.12}) it follows that
\begin{equation}
\Omega_{0}(Y_{1}^{(k)},Y_{1}^{(m)})=0\Longrightarrow\Pi_{0}(dF^{(k)}%
,dF^{(m)})=0\Longrightarrow\lbrack X^{(k)},X^{(m)}]=0. \label{3.15}%
\end{equation}
Now we show that the Schouten bracket of $\Pi_{1}$ is zero. As $\Pi_{0}%
\Omega_{1}\Pi_{0}$ is Poisson (it follows from compatibility of $\Omega_{0}$
and $\Omega_{1}$), we have
\[
\lbrack\Pi_{1},\Pi_{1}]_{S}=2%
{\textstyle\sum\nolimits_{k}}
[\Pi_{0}\Omega_{1}\Pi_{0},X^{(k)}\wedge Y_{0}^{(k)}]_{S}+%
{\textstyle\sum\nolimits_{k,m}}
[X^{(k)}\wedge Y_{0}^{(k)},X^{(m)}\wedge Y_{0}^{(m)}]_{S},
\]%
\[
\lbrack\Pi_{0}\Omega_{1}\Pi_{0},X^{(k)}\wedge Y_{0}^{(k)}]_{S}=Y_{0}%
^{(k)}\wedge\Pi_{0}d(\Omega_{1}X^{(k)})\Pi_{0}-X^{(k)}\wedge\Pi_{0}%
d(\Omega_{1}Y_{0}^{(k)})\Pi_{0},
\]%
\[
\lbrack X^{(k)}\wedge Y_{0}^{(k)},X^{(m)}\wedge Y_{0}^{(m)}]_{S}%
=2X^{(k)}\wedge Y_{0}^{(m)}\wedge\lbrack Y_{0}^{(k)},X^{(m)}].
\]
In last equality we used the fact that $[Y_{0}^{(k)},Y_{0}^{(m)}]=0$ and
relation (\ref{3.15}). Now,
\begin{align*}
\lbrack Y_{0}^{(k)},X^{(m)}]  &  =[Y_{0}^{(k)},\Pi_{0}\Omega_{1}Y_{0}%
^{(m)}]=L_{Y_{0}^{(k)}}(\Pi_{0}\Omega_{1})Y_{0}^{(m)}=\Pi_{0}(L_{Y_{0}^{(k)}%
}\Omega_{1})Y_{0}^{(m)}\\
&  =\Pi_{0}d(\Omega_{1}\mu Y_{0}^{(k)})Y_{0}^{(m)}=\Pi_{0}(d\beta^{(k)}%
)Y_{0}^{(m)}\\
&  =\Pi_{0}\left[  -%
{\textstyle\sum\nolimits_{i}}
d(Y_{0}^{(k)}(F^{(i)}))\wedge dH_{0}^{(i)}]\right]  Y_{0}^{(m)}\\
&  =-\Pi_{0}d(Y_{0}^{(k)}(F^{(m)})).
\end{align*}
From
\begin{equation}
Y_{1}^{(k)}=X^{(k)}+%
{\textstyle\sum\nolimits_{m}}
Y_{1}^{(k)}(H_{0}^{(m)})Y_{0}^{(m)}\nonumber
\end{equation}
we have
\[
\Omega_{1}Y_{1}^{(k)}=\Omega_{1}X^{(k)}+%
{\textstyle\sum\nolimits_{m}}
Y_{1}^{(k)}(H_{0}^{(m)})\Omega_{1}Y_{0}^{(m)}%
\]
and
\begin{equation}
\Omega_{1}X^{(k)}=dG^{(k)}-%
{\textstyle\sum\nolimits_{i}}
Y_{0}^{(k)}(G^{(i)})dH_{0}^{(i)}-%
{\textstyle\sum\nolimits_{i}}
Y_{1}^{(k)}(H_{0}^{(i)})dF^{(i)}+%
{\textstyle\sum\nolimits_{i,j}}
Y_{1}^{(k)}(H_{0}^{(i)})Y_{0}^{(i)}(F^{(j)})dH_{0}^{(j)}.\nonumber
\end{equation}
Hence,
\begin{align*}
\Pi_{0}d(\Omega_{1}X^{(k)})\Pi_{0}  &  =-\Pi_{0}\left[
{\textstyle\sum\nolimits_{m}}
d(Y_{1}^{(k)}(H_{0}^{(m)}))\wedge dF^{(m)}\right]  \Pi_{0}\\
&  =%
{\textstyle\sum\nolimits_{m}}
\Pi_{0}d(Y_{1}^{(k)}(H_{0}^{(m)}))\wedge X^{(m)}%
\end{align*}
and then $[\Pi_{1},\Pi_{1}]_{S}=0$ under condition (\ref{3.14}).$\Box$

\section{Bi-presymplectic chains}

Now we are ready to investigate main properties of bi-presymplectic chains.

\begin{theorem}
\label{theorem3} Assume we have a pair of presymplectic forms $(\Omega
_{0},\Omega_{1})$, d-compatible with respect to some $\Pi_{0}$ dual to
$\Omega_{0},\,$\ both of rank $2n$ and co-rank $r$ on $\mathcal{O}\subset \mathcal{M}$. Assume further, that they
form\ bi-presymplectic chains of one-forms
\begin{equation}
\beta_{i}^{(k)}=\Omega_{0}Y_{i}^{(k)}=\Omega_{1}Y_{i-1}^{(k)},\quad
i=1,2,\dots,n_{k} \label{beta}%
\end{equation}
where $k=1,...,r,$ $n_{1}+...+n_{r}=n$ and each chain starts with a kernel
vector field $Y_{0}^{(k)}$ of $\Omega_{0}$ and terminates with a kernel vector
field $Y_{n_{k}}^{(k)}$ of $\Omega_{1}$. Then \vspace{0.4cm}\newline(i)
\begin{equation}
\Omega_{0}(Y_{i}^{(k)},Y_{j}^{(m)})=\Omega_{1}(Y_{i}^{(k)},Y_{j}^{(m)})=0,
\label{t3e1}%
\end{equation}
for $k,m=1,...,r,$ $i=1,2,\dots,n_{k},$ $j=1,2,\dots,n_{m}.$

Moreover, let us assume that
\begin{equation}
X_{i}^{(k)}=\Pi_{0}\beta_{i}^{(k)}=\Pi_{0}dH_{i}^{(k)}, \label{x}%
\end{equation}
for $k=1,...,r,$ $i=1,2,\dots,n_{k},$ which implies%
\begin{equation}
\beta_{i}^{(k)}=dH_{i}^{(k)}-%
{\textstyle\sum\nolimits_{m}}
Y_{0}^{(m)}(H_{i}^{(k)})dH_{0}^{(m)}, \label{4.1}%
\end{equation}%
\begin{equation}
Y_{i}^{(k)}=X_{i}^{(k)}+%
{\textstyle\sum\nolimits_{m}}
Y_{i}^{(k)}(H_{0}^{(m)})Y_{0}^{(m)}, \label{4.2}%
\end{equation}
where $\Pi_{0}dH_{0}=0$. Then,\vspace{0.4cm}\newline(ii)
\begin{equation}
\Pi_{0}(dH_{i}^{(k)},dH_{j}^{(m)})=0,\quad\lbrack X_{i}^{(k)},X_{j}^{(m)}]=0
\label{4.3}%
\end{equation}
and equations (\ref{beta}) define a Liouville integrable system.\vspace
{0.4cm}\newline

Additionally, if
\begin{equation}
Y_{0}^{(k)}(H_{1}^{(m)})=Y_{0}^{(m)}(H_{1}^{(k)}) \label{4.4}%
\end{equation}
and
\begin{equation}
Y_{0}^{(k)}(H_{i}^{(m)})=Y_{i}^{(m)}(H_{0}^{(k)}), \label{4.5}%
\end{equation}
where $k,m=1,...,r,$ $i=1,2,\dots,n_{m}$,\ then\vspace{0.4cm}\newline(iii)
vector fields $X_{i}^{(k)}$ (\ref{x}) form bi-Hamiltonian chains
\begin{equation}
X_{i}^{(k)}=\Pi_{0}dH_{i}^{(k)}=\Pi_{1}dH_{i-1}^{(k)},\quad i=1,2,\dots,n
\label{4.6}%
\end{equation}
where
\begin{equation}
\Pi_{1}=\Pi_{0}\Omega_{1}\Pi_{0}+%
{\textstyle\sum\nolimits_{m}}
X_{1}^{(m)}\wedge Y_{0}^{(m)}, \label{4.7}%
\end{equation}
$k,m=1,...,r,$ $i=1,2,\dots,n_{k}$ and $n_{1}+...+n_{r}=n$. The chain starts
with $H_{0}^{(k)}$, a Casimir of $\Pi_{0}$, and terminates with $H_{n_{k}%
}^{(k)}$, a Casimir of $\Pi_{1}$. Moreover the Poisson pair $(\Pi_{0},\Pi
_{1})$ is d-compatible with respect to $\Omega_{0}$.
\end{theorem}

\textbf{Proof.}\newline(i) From (\ref{beta}) we have
\[%
\begin{array}
[c]{l}%
\Omega_{0}(Y_{i}^{(k)},Y_{j}^{(m)})=\Omega_{0}(Y_{i-1}^{(k)},Y_{j+1}^{(m)}),\\
\Omega_{1}(Y_{i}^{(k)},Y_{j}^{(m)})=\Omega_{1}(Y_{i+1}^{(k)},Y_{j-1}^{(m)})\\
\Omega_{0}(Y_{i}^{(k)},Y_{j}^{(m)})=\Omega_{1}(Y_{i-1}^{(k)},Y_{j}^{(m)}).
\end{array}
\]
Then, (\ref{t3e1}) follows from
\begin{equation}
\Omega_{0}(Y_{0}^{(k)},Y_{i}^{(m)})=0,\quad\Omega_{1}(Y_{n_{k}}^{(k)}%
,Y_{i}^{(m)})=0.\nonumber
\end{equation}
(ii) From properties of dual pair $(\Pi_{0},\Omega_{0})$, if $X_{i}^{(k)}%
=\Pi_{0}dH_{i}^{(k)}$ then
\begin{equation}
\Pi_{0}(dH_{i}^{(k)},dH_{j}^{(m)})=\Omega_{0}(X_{i}^{(k)},X_{j}^{(m)}%
).\nonumber
\end{equation}
On the other hand as $X_{i}^{(k)}=Y_{i}^{(k)}+%
{\textstyle\sum\nolimits_{m}}
\alpha_{m}^{(k)}Y_{0}^{(m)},$ where $\alpha_{m}^{(k)}$ are an appropriate
functions, it follows that
\begin{equation}
\Omega_{0}(X_{i}^{(k)},X_{j}^{(m)})=\Omega_{0}(Y_{i}^{(k)},Y_{j}%
^{(m)}).\nonumber
\end{equation}
(iii) We have
\begin{align*}
X_{i}^{(k)}  &  =\Pi_{0}dH_{i}^{(k)}\\
&  =\Pi_{0}\Omega_{1}Y_{i-1}^{(k)}=\Pi_{0}\Omega_{1}(X_{i-1}^{(k)}+%
{\textstyle\sum\nolimits_{m}}
Y_{i-1}^{(k)}(H_{0}^{(m)})Y_{0}^{(m)})\\
&  =\Pi_{0}\Omega_{1}\Pi_{0}dH_{i-1}^{(k)}+%
{\textstyle\sum\nolimits_{m}}
Y_{i-1}^{(k)}(H_{0}^{(m)})X_{1}^{(m)})\\
&  \overset{(\ref{4.5})}{=}(\Pi_{0}\Omega_{1}\Pi_{0}+%
{\textstyle\sum\nolimits_{m}}
X_{1}^{(m)}\wedge Y_{0}^{(m)})dH_{i-1}^{(k)}\\
&  =\Pi_{1}dH_{i-1}^{(k)}.
\end{align*}
Moreover, $\Pi_{0}$ and $\Pi_{1}$ are d-compatible Poisson tensors provided
that (\ref{4.4}) is fulfilled. We also have
\begin{align*}
\Pi_{1}dH_{n_{k}}^{(k)}  &  =(\Pi_{0}\Omega_{1}\Pi_{0}+%
{\textstyle\sum\nolimits_{m}}
X_{1}^{(m)}\wedge Y_{0}^{(m)})dH_{n_{k}}^{(k)}=\Pi_{0}\Omega_{1}X_{n_{k}%
}^{(k)}+%
{\textstyle\sum\nolimits_{m}}
Y_{0}^{(m)}(H_{n_{k}}^{(k)})X_{1}^{(m)}\\
&  \overset{(\ref{4.2})}{=}\,\Pi_{0}\Omega_{1}(Y_{n_{k}}^{(k)}-%
{\textstyle\sum\nolimits_{m}}
Y_{n_{k}}^{(k)}(H_{0}^{(m)})Y_{0}^{(m)})+%
{\textstyle\sum\nolimits_{m}}
Y_{0}^{(m)}(H_{n_{k}}^{(k)})X_{1}^{(m)}\\
&  =-%
{\textstyle\sum\nolimits_{m}}
Y_{n_{k}}^{(k)}(H_{0}^{(m)})X_{1}^{(m)}+%
{\textstyle\sum\nolimits_{m}}
Y_{0}^{(m)}(H_{n_{k}}^{(k)})X_{1}^{(m)}\overset{(\ref{4.5})}{=}0.
\end{align*}

$\Box$

Notice, that in a special case, when
\begin{equation}
Y_{0}^{(k)}(H_{i}^{(m)})=0, \label{4.8}%
\end{equation}
for all admissible values of $k,m$ and $i$, chains (\ref{beta}) are
bi-inverse-Hamiltonian as $\beta_{i}^{(k)}=dH_{i}^{(k)}$. Obviouslu
$X_{i}^{(k)}$ are not bi-Hamiltonian until $Y_{i}^{(k)}(H_{0}^{(m)})\neq0$.

Finally we show that arbitrary Liouville integrable system which has a
bi-presymplectic representation on $(2n+r)$-dimensional phase space, has also
quasi-bi-Hamiltonian representation on any symplectic leaf of its Hamiltonian
structure $\Pi_{0}$. Actually, from (\ref{beta}), (\ref{x}) and (\ref{4.1})
follows that
\begin{align*}
\Pi_{0}dH_{i}^{(k)}  &  =\Pi_{0}\Omega_{1}\left(  Y_{i-1}^{(k)}+%
{\textstyle\sum\nolimits_{m}}
Y_{0}^{(m)}(H_{i}^{(k)})dH_{0}^{(m)}\right) \\
&  =\Pi_{0}\left[  \Omega_{1}X_{i-1}^{(k)}+%
{\textstyle\sum\nolimits_{m}}
\left(  Y_{i-1}^{(k)}(H_{0}^{(m)})\Omega_{1}Y_{0}^{(m)}+Y_{0}^{(m)}%
(H_{i}^{(k)})dH_{0}^{(m)}\right)  \right] \\
&  =\Pi_{0}\Omega_{1}\Pi_{0}dH_{i-1}^{(k)}+%
{\textstyle\sum\nolimits_{m}}
Y_{i-1}^{(k)}(H_{0}^{(m)})\Pi_{0}dH_{1}^{(m)},
\end{align*}
hence on $(2n+r)$-dimensional phase space we have quasi-bi-Hamiltonian
representation
\begin{equation}
\Pi_{1}dH_{i-1}^{(k)}=\Pi_{0}dH_{i}^{(k)}+\sum_{m=1}^{r}F_{i-1}^{(k,m)}\Pi
_{0}dH_{1}^{(m)}, \label{4.9}%
\end{equation}
where
\[
\Pi_{1}=\Pi_{0}\Omega_{1}\Pi_{0},\ \ \ \ \ \ F_{i}^{(k,m)}=-Y_{i}^{(k)}%
(H_{0}^{(m)}).
\]

Notice that both Poisson structures $\Pi_{0}$ and $\Pi_{1}=\Pi_{0}\Omega
_{1}\Pi_{0}$ share the same Casimirs $H_{0}^{(k)}$, so the
quasi-bi-Hamiltonian dynamics can be restricted immediately to any common leaf
of dimension $2n$
\begin{equation}
\pi_{1}dh_{i-1}^{(k)}=\pi_{0}dh_{i}^{(k)}+\sum_{m=1}^{r}F_{i-1}^{(k,m)}\pi
_{0}dh_{1}^{(m)},\qquad i=1,...,n, \label{qh}%
\end{equation}
where $\pi_{i}$ and $h_{i}^{(k)}$are restrictions of $\Pi_{i}$ and
$H_{i}^{(k)}$, respectively. Hence we deal with a St{\"{a}}ckel system whose
separation coordinates are eigenvalues of the recursion operator $N=\pi_{1}%
\pi_{0}^{-1}$ \cite{mag}, provided that $N$ has $n$ distinct and functionally
independent eigenvalues at any point of $\mathcal{O}\subset \mathcal{M}$, i.e. we are in a generic case. We
will come back to separable systems in next sections.

The advantage of bi-presymplectic representation of Liouville integrable
system, when compared to bi-Hamiltonian ones, is that the existence of the
first guarantees that the system is separable and the construction of
separation coordinates is purely algorithmic (in a generic case), while the
bi-Hamiltonian representation does not guarantee the existence of
quasi-bi-Hamiltonian representation and hence separability of the system in
question. Moreover, the projection of the second Poisson structure onto the
symplectic foliation of the first one, in order to construct a
quasi-bi-Hamiltonian representation, necessary for separability, is far from
being trivial non-algorithmic procedure \cite{m4}.

\section{Separable St\"{a}ckel systems}

Consider a Liouville integrable system on a $2n$-dimensional phase space $M,$
i.e. a set of $h_{i}\in C(M),$ $i=1,...,n$ which are in involution with
respect to some Poisson tensor $\pi_{0}$, and related $n$ Hamiltonian dynamic
systems
\begin{equation}
u_{t_{i}}=\pi\,dh_{i}=x_{h_{i}},\qquad i=1,\dots,n, \label{1}%
\end{equation}
where $u\in M$ and $x_{h_{i}}$ are respective Hamiltonian vector fields.

The Hamilton-Jacobi (HJ) method for solving (\ref{1}) essentially amounts to
the linearization of the latter via a canonical transformation (when $u\in \mathcal{O}\subset \mathcal{M}$
is some local canonical chart)
\begin{equation}
u=(q,p)\rightarrow\ (b,a),\quad a_{i}=h_{i},\qquad i=1,\dots,n. \label{2}%
\end{equation}
In order to find the conjugate coordinates $b^{i}$ it is necessary to
construct a generating function $W(q,a)$ of the transformation (\ref{2}) such
that
\[
b^{j}=\tfrac{\partial W}{\partial a_{j}},\quad p_{j}=\tfrac{\partial
W}{\partial q^{j}}.
\]

The function $W(q,a)$ is a complete integral of the associated
\emph{Hamilton-Jacobi equations}
\begin{equation}
h_{i}\left(  q^{1},\dots,q^{n},\tfrac{\partial W}{\partial q^{1}},\dots
,\tfrac{\partial W}{\partial q^{n}}\right)  =a_{i},\qquad i=1,\dots,n.
\label{3}%
\end{equation}
In the $(b,a)$ representation the $t_{i}$-dynamics is trivial:
\[
(a_{j})_{t_{i}}=0,\quad(b^{j})_{t_{i}}=\delta_{ij},
\]
whence
\begin{equation}
b^{j}(q,a)=\tfrac{\partial W}{\partial a_{j}}=t_{j}+\gamma_{j},\qquad
j=1,\dots,n, \label{4}%
\end{equation}
where $\gamma_{j}$ are arbitrary constants.

Equations (\ref{4}) provide implicit solutions for (\ref{1}). Solving them for
$q_{j}$ is known as the \emph{inverse Jacobi problem}. The reconstruction in
explicit form of trajectories $q^{j}=q^{j}(t_{i})$ is itself a highly
nontrivial problem from algebraic geometry.

To solve the system (\ref{3}) for $W$ in a given local coordinate system is a
hopeless task, as (\ref{3}) is a system of nonlinear coupled partial
differential equations. In essence, the only hitherto known way of overcoming
this difficulty is to find distinguished canonical coordinates, denoted here
by $(\lambda,\mu)$, for which there exist $n$ relations
\begin{equation}
\varphi_{i}(\lambda^{i},\mu_{i};a_{1},\dots,a_{n})=0,\qquad i=1,\dots
,n,\ a_{i}\in\mathbb{R}\mathbf{,}\quad\det\left[  \tfrac{\partial\varphi_{i}%
}{\partial a_{j}}\right]  \neq0, \label{5}%
\end{equation}
such that each of these relations involves only a single pair of canonical
coordinates \cite{sk1}. The determinant condition in (\ref{5}) means that we
can solve the equations (\ref{5}) for $a_{i}$ and express $a_{i}$ in the form
$a_{i}=h_{i}(\lambda,\mu)$, $i=1,\ldots,n$.

If the functions $W_{i}(\lambda^{i},a)~$\ are solutions of a system of $n$
\emph{decoupled} ODEs obtained from (\ref{5}) by substituting $\mu_{i}%
=\tfrac{dW_{i}(\lambda^{i},a)}{d\lambda^{i}}$%
\begin{equation}
\varphi_{i}\left(  \lambda^{i},\mu_{i}=\tfrac{dW_{i}(\lambda^{i},a)}%
{d\lambda^{i}},a_{1},\ldots,a_{n}\right)  =0,\quad i=1,\dots,n, \label{SRdiff}%
\end{equation}
then the function
\[
W(\lambda,a)=\sum\nolimits_{i=1}^{n}W_{i}(\lambda^{i},a)
\]
is an additively separable solution of \emph{all} the equations (\ref{SRdiff}%
), and \emph{simultaneously} it is a solution of all Hamilton-Jacobi equations
(\ref{3}). The distinguished coordinates $(\lambda,\mu)$ for which the
original Hamilton-Jacobi equations (\ref{3}) are equivalent to a set of
separation relations (\ref{SRdiff}) are called the \emph{separation
coordinates}.

In what follows we restrict ourselves to considering a special case of
(\ref{5}) when all separation relations are affine in $h_{i}$:
\begin{equation}
\sum_{k=1}^{n}S_{i}^{k}(\lambda^{i},\mu_{i})h_{k}=\psi_{i}(\lambda^{i},\mu
_{i}),\qquad i=1,\dots,n, \label{6}%
\end{equation}
where $S_{i}^{k}$ and $\psi_{i}$ are arbitrary smooth functions of their
arguments. The relations (\ref{6}) are called the generalized
\emph{St\"{a}ckel separation relations} and the related dynamical systems are
called the \emph{St\"{a}ckel separable} ones. The matrix $S=(S_{i}^{k})$ will
be called a \emph{generalized St\"{a}ckel matrix}. To recover the explicit
St\"{a}ckel form of the Hamiltonians it suffices to solve the linear system
(\ref{6}) with respect to $h_{i}$.

If in (\ref{6}) we further have $S_{i}^{k}(\lambda^{i},\mu_{i})=S^{k}%
(\lambda^{i},\mu_{i})$ and $\psi_{i}(\lambda^{i},\mu_{i})=\psi(\lambda^{i}%
,\mu_{i})$ then the separation conditions can be represented by $n$ copies of
the curve
\begin{equation}
\sum_{k=1}^{n}S^{k}(\lambda,\mu)h_{k}=\psi(\lambda,\mu) \label{7}%
\end{equation}
in $(\lambda,\mu)$ plane, called a \emph{separation curve}. The copies in
question are obtained by setting $\lambda=\lambda^{i}$ and $\mu=\mu_{i}$ for
$i=1,\dots,n$.

For further convenience, let us collect the terms from the l.h.s. of (\ref{6})
as follows:
\begin{equation}
\sum_{k=1}^{r}\varphi_{i}^{k}(\lambda^{i},\mu_{i})h^{(k)}(\lambda^{i}%
)=\psi_{i}(\lambda^{i},\mu_{i}),\qquad i=1,\dots,n, \label{9}%
\end{equation}
where
\[
h^{(k)}(\lambda)=\sum_{i=1}^{n_{k}}\lambda^{n_{k}-i}h_{i}^{(k)},\qquad
n_{1}+\dots+n_{r}=n
\]
and impose the normalization $\varphi_{i}^{r}(\lambda^{i},\mu_{i})=1$.

Some informations about the classification of St\"{a}ckel systems (\ref{9})
the reader can find in \cite{bb1}.

\section{Bi-inverse-Hamiltonian representation of St\"{a}ckel systems}

As recently proved in \cite{bb1}, the St\"{a}ckel Hamiltonians defined by
separation relations (\ref{9}) admit on $M$ the following quasi bi-Hamiltonian
chains in $(\lambda,\mu)$ representation
\begin{equation}
\pi_{1}dh_{i}^{(k)}=\pi_{0}\,dh_{i+1}^{(k)}+\sum_{l=1}^{r}F_{i}^{(k,l)}%
\,\pi_{0}\,dh_{1}^{(l)},\ \ \ h_{n_{k}+1}^{(k)}=0,\ \ k=1,\dots,r,~~i=1,\dots
,n_{k},\label{0.7}%
\end{equation}
where $\pi_{0}$ is a canonical Poisson tensor
\[
\pi_{0}=\sum_{i}\frac{\partial}{\partial\lambda^{i}}\wedge\frac{\partial
}{\partial\mu_{i}},
\]
$\pi_{1}$ is a noncanonical Poisson tensor of the form
\[
\pi_{1}=\sum_{i}\lambda^{i}\frac{\partial}{\partial\lambda^{i}}\wedge
\frac{\partial}{\partial\mu_{i}},
\]
compatible with $\pi_{0}$, and the \emph{expantion cefficients} $F_{i}%
^{(k,l)}$ are solutions of the set of linear algebraic equations
\begin{equation}
\sum_{k=1}^{r}\varphi_{j}^{k}(\lambda^{j},\mu_{j})F^{(k,l)}(\lambda
^{j})=\varphi_{j}^{l}(\lambda^{j},\mu_{j})(\lambda^{j})^{n_{l}},\qquad
j=1,\dots,n,\ \ \ \ l=1,...,r,\label{0.8}%
\end{equation}
where
\[
F^{(k,l)}(\lambda)=\sum_{i=1}^{n_{k}}\lambda^{n_{k}-i}F_{i}%
^{(k,l)},\qquad n_{1}+\dots+n_{r}=n.
\]

Let us consider the following symplectic forms on $M$
\[
\omega_{0}=-\sum_{i}d\lambda^{i}\wedge d\mu_{i},\ \ \ \ \ \ \omega_{1}%
=-\sum_{i}\lambda^{i}d\lambda^{i}\wedge d\mu_{i}.
\]
Observe that $(\pi_{0},\omega_{0})$ constituts non degenerate dual
implectic-symplectic pair as $\omega_{0}=\pi_{0}^{-1}$, $\pi_{0}$ and $\pi
_{1}=\pi_{0}\omega_{1}\pi_{0}$ are d-compatible with respect to $\omega_{0}$
and $\omega_{0}$ and $\omega_{1}=\omega_{0}\pi_{1}\omega_{0}$ are d-compatible
with respect to $\pi_{0}$. Besides, quasi bi-Hamiltonian chains (\ref{0.7})
have equivalent quasi bi-inverse-Hamiltonian representations
\begin{equation}
\omega_{1}x_{i}^{(k)}=\omega_{0}\,x_{i+1}^{(k)}+\sum_{l=1}^{r}F_{i}%
^{(k,l)}\,\omega_{0}\,x_{1}^{(l)},\ \ \ \ x_{n_{k}+1}^{(k)}=0,\ \ \ k=1,\dots
,r,\quad i=1,\dots,n_{k},\label{0.9}%
\end{equation}
where
\[
x_{i}^{(k)}=\pi_{0}dh_{i}^{(k)},\ \ \ \ \ \ \ \ dh_{i}^{(k)}=\omega_{0}%
x_{i}^{(k)}.
\]

Let us lift the whole construction to the extended phase space $M\rightarrow
\mathcal{M}:$ $(\lambda,\mu)\rightarrow(\lambda,\mu,c)$, where $\dim
\mathcal{M}=2n+r.$ Then, on $\mathcal{M}:$ $\omega_{0}\rightarrow\Omega_{0},$
$\pi_{0}\rightarrow\Pi_{0},$ both degenerated,  where
\[
\ker\Omega_{0}=Sp\{Y_{0}^{(k)}\},\ \ k=1,...,r,\ \ \ Y_{0}^{(k)}
=\frac{\partial}{\partial c_{k}},\ \ \ \Omega_{0}Y_{0}^{(k)}=0
\]
and
\[
\ker\Pi_{0}=Sp\{dc_{k}\},\ \ k=1,...,r,\ \ \ \ \ \Pi_{0}dc_{k}=0,\ \ \ \ Y_{0}%
^{(k)}(c_{j})=\delta_{j}^k.
\]
Obviously, $(\Pi_{0},\Omega_{0})$ is a dual Poisson-presymplectic pair on
$\mathcal{M}$. In the same fasion we lift
\[
\omega_{1}\rightarrow\Omega_{1D},\ \ \ \ \pi_{1}\rightarrow\Pi_{1D}%
,\ \ \ \ x_{i}^{(k)}\rightarrow X_{i}^{(k)},
\]
where $\ker\Omega_{1D}=\ker\Omega_{0}$ and $\ker\Pi_{1D}=\ker\Pi_{0}.$ On
$\mathcal{M}$ quasi bi-inverse-Hamiltonian chains (\ref{0.9}) take the form
\begin{equation}
\Omega_{1D}X_{i}^{(k)}=\Omega_{0}\,X_{i+1}^{(k)}+\sum_{l=1}^{r}F_{i}%
^{(k,l)}\,\Omega_{0}\,X_{1}^{(l)},\ \ \ \ k=1,\dots,r,\quad i=1,\dots
,n_{k}.\label{0.10}%
\end{equation}

Let us define the following presymplectic two-form
\begin{equation}
\Omega_{1}=\Omega_{1D}+\sum_{k=1}^{r}dh_{1}^{(k)}\wedge dc_{k} \label{0.11}%
\end{equation}
and the set of vector fields
\begin{equation}
Y_{i}^{(k)}=X_{i}^{(k)}-\sum_{l=1}^{r}F_{i}^{(k,l)}Y_{0}^{(l)}. \label{0.12}%
\end{equation}
Then, we have
\begin{align*}
\Omega_{0}Y_{i+1}^{(k)}  &  =dh_{i+1}^{(k)}\\
&  =\Omega_{0}X_{i+1}^{(k)}=\Omega_{1D}X_{i}^{(k)}-\sum_{l=1}^{r}F_{i}%
^{(k,l)}\,\Omega_{0}\,X_{1}^{(l)}\\
&  =(\Omega_{1}-\sum_{l=1}^{r}dh_{1}^{(l)}\wedge dc_{l})(Y_{i}^{(k)}%
+\sum_{l=1}^{r}F_{i}^{(k,l)}Y_{0}^{(l)})-\sum_{l=1}^{r}F_{i}^{(k,l)}%
dh_{1}^{(l)}\\
&  =\Omega_{1}Y_{i}^{(k)}+\sum_{l=1}^{r}F_{i}^{(k,l)}\Omega_{1}Y_{0}%
^{(l)}-\sum_{l=1}^{r}Y_{i}^{(k)}(c_{l})dh_{1}^{(l)}+\sum_{l=1}^{r}Y_{i}%
^{(k)}(h_{1}^{(l)})dc_{l}\\
&  -\sum_{l=1}^{r}F_{i}^{(k,l)}dh_{1}^{(l)}+\sum_{l,m=1}^{r}F_{i}^{(k,m)}%
Y_{0}^{(m)}(h_{1}^{(l)})dc_{l}-\sum_{l=1}^{r}F_{i}^{(k,l)}dh_{1}^{(l)}\\
&  =\Omega_{1}Y_{i}^{(k)},
\end{align*}
as
\[
\Omega_{1}Y_{0}^{(l)}=\sum_{k=1}^{r}(dh_{1}^{(k)}\wedge dc_{k})Y_{0}%
^{(l)}=dh_{1}^{(l)},
\]%
\[
Y_{i}^{(k)}(h_{1}^{(l)})=0,\ \ \ Y_{i}^{(k)}(c_{l})=-F_{i}^{(k,l)}%
,\ \ \ Y_{0}^{(m)}(c_{k})=\delta_{mk.}%
\]
Hence, on $\mathcal{M}$, differentials $dh_{i}^{(k)}$ form a
bi-inverse-Hamiltonian hierarchies
\begin{equation}
\Omega_{0}Y_{i+1}^{(k)}=dh_{i+1}^{(k)}=\Omega_{1}Y_{i}^{(k)},\quad
i=0,1,2,\dots,n_{k},\ \ \ k=1,...,r, \label{0.13}%
\end{equation}
which starts with a kernel vector field
$Y_{0}^{(k)}$ of $\Omega_{0}$ and terminates with a kernel vector field
$Y_{n_{k}}^{(k)}$ of $\Omega_{1}$. Indeed
\begin{align*}
\Omega_{1}Y_{n_{k}}^{(k)}  &  =(\Omega_{1D}+\sum_{m=1}^{r}dh_{1}^{(m)}\wedge
dc_{m})(X_{n_{k}}^{(k)}-\sum_{m=1}^{r}F_{n_{k}}^{(k,m)}Y_{0}^{(m)})\\
&  =\sum_{m=1}^{r}F_{n_{k}}^{(k,m)}dh_{1}^{(m)}-\sum_{m=1}^{r}F_{n_{k}%
}^{(k,m)}dh_{1}^{(m)}=0.
\end{align*}
Moreover, $\Omega_{0}$ and $\Omega_{1}$ are d-compatible with respect to
$\Pi_{0}$, as
\[
\Pi_{0}\Omega_{1}\Pi_{0}=\Pi_{0}\Omega_{1D}\Pi_{0}=\Pi_{1D}%
\]
which is Poisson. According to theorem \ref{theorem3} vector fields
$X_{i}^{(k)}$ are not bi-Hamiltonian as $Y_{i}^{(k)}(h_{0}^{(l)}%
)=-F_{i}^{(k,l)}\neq0.$

In order to construct on $\mathcal{M}$ bi-Hamiltonian representation of considered St\"{a}ckel systems,
one has to extend the original Hamiltonians
\begin{equation}
h_{i}^{(k)}\rightarrow H_{i}^{(k)}=h_{i}^{(k)}-\sum_{l=1}^{r}F_{i}%
^{(k,l)}c_{l},\,\,\,\,\,\,\,i=1,...,n. \label{0.14}%
\end{equation}
Then, on $\mathcal{M}$, vector fields
\begin{equation}
K_{i}^{(k)}=X_{i}^{(k)}-\Pi_{0}d(\sum_{l=1}^{r}F_{i}^{(k,l)}c_{l})
\label{0.15}%
\end{equation}
form a bi-Hamiltonian chains
\begin{equation}
\Pi_{0}dH_{i+1}^{(k)}=K_{i+1}^{(k)}=\Pi_{1}dH_{i}^{(k)},\ \ \ i=0,1,\dots
,n_{k},\ \ \ k=1,...,r, \label{0.16}%
\end{equation}
where
\begin{equation}
\Pi_{1}=\Pi_{1D}+\sum_{m=1}^{r}K_{1}^{(m)}\wedge Y_{0}^{(m)} \label{0.17}%
\end{equation}
is a Poisson tensor compatible with $\Pi_{0}$ one. Each chain starts with the
Casimir of $\Pi_{0}$, i.e. $H_{0}^{(k)}=c_{k}$, and terminates with the
Casimir of $\Pi_{1}$, i.e. $H_{n_{k}}^{(k)}$. The details of the construction
the reader finds in \cite{bb1}. Poisson tensors $\Pi_{0}$ and $\Pi_{1} $ are
d-compatible with respect to $\Omega_{0}$ as
\[
\Omega_{0}\Pi_{1}\Omega_{0}=\Omega_{0}\Pi_{1D}\Omega_{0}=\Omega_{1D}%
\]
is closed. As was proved in \cite{b}, bi-Hamiltonian chains (\ref{0.16}) have
no bi-presymplectic counterparts as the conditions (\ref{4.4}) are not
satisfied (see also theorem \ref{theorem2}). Indeed
\[
Y_{0}^{(k)}(H_{1}^{(m)})=-F_{1}^{(m,k)}\neq-F_{1}^{(k,m)}=Y_{0}^{(m)}%
(H_{1}^{(k)}).
\]
The only exception is the case of co-rank one ($r=1$), as then (\ref{4.4}) is
trivially fulfilled.

\section{Examples}

Here we illustrate the presented theory with three examples of separable
systems, each of three degrees of freedom. Two of them are classical
St\"{a}ckel systems with separation relations being quadratic in momenta,
while the third example has separation relations cubic in momenta.\newline%
\textbf{Example 1.}\newline Consider the separation relations on a
six-dimensional phase space $M$ given by the following bare (potential-free)
separation curve
\[
h_{1}\lambda^{2}+h_{2}\lambda+h_{3}=\tfrac{1}{2}\mu^{2}.\label{7.21a}%
\]
This curve corresponds to geodesic motion for a classical St\"{a}ckel system
(of Benenti type \cite{mac2005}). As in this example $k=1$, we use the
notation $h_{i}^{(1)}\equiv h_{i}$. The transformation $(\lambda
,\mu)\rightarrow(q,p)$ to the flat coordinates of associated metric follows
from the point transformation
\begin{eqnarray*}
\sigma_1(q)&=&q_{1}=-\lambda ^{1}-\lambda ^{2}-\lambda ^{3},\\ \nonumber
\sigma_2(q)&=&\tfrac{1}{4}q_{1}^{2}+%
q_{2}=\lambda ^{1}\lambda ^{2}+\lambda ^{1}\lambda ^{3}+\lambda
^{2}\lambda ^{3},\\ \nonumber
\sigma_3(q)&=&\tfrac{1}{2}q_{1}q_{2}+q_{3}=-\lambda
^{1}\lambda ^{2}\lambda ^{3}. \label{qp}
\end{eqnarray*}
In the flat coordinates the Hamiltonians take the form
\begin{align*}
E_{1}& =p_{1}p_{3}+\tfrac{1}{2}p_{2}^{2}, \nonumber \\
E_{2}& =p_1p_2+\tfrac{1}{2}q_{1}p_{2}^{2}+\tfrac{1}{2}q_{1}p_{1}p_3-\tfrac{1}{2}%
q_{2}p_{2}p_{3}-\tfrac{1}{2}q_{3}p_{3}^2, \nonumber  \\
E_{3}& =\tfrac{1}{2}p_{1}^{2}+\tfrac{1}{8}q_{1}^{2}p_{2}^{2}+%
\tfrac{1}{8}q_2^2p_{3}^{2}+\tfrac{1}{2}q_{1}p_{1}p_{2}+\tfrac{1}{2}%
q_{2}p_{1}p_{3} \label{E} \\
&-(\tfrac{1}{4}q_{1}q_{2}+q_{3})p_{2}p_{3}, \nonumber
\end{align*}
and admit a quasi bi-inverse-Hamiltonian representation (\ref{0.9})
\begin{align*}
\omega_{1}x_{1} &  =\omega_{0}x_{2}+F_{1}\omega_{0}x_{1},\\
\omega_{1}x_{2} &  =\omega_{0}x_{3}+F_{2}\omega_{0}x_{1},\\
\omega_{1}x_{3} &  =F_{3}\omega_{0}x_{1},
\end{align*}
with the operators $\omega_{0}$ and $\omega_{1}$ of the form
\begin{equation}
\omega_{0}=\pi_{0}^{-1}=\left(
\begin{array}
[c]{cc}%
0 & -I_{3}\\
I_{3} & 0
\end{array}
\right)  ,\label{p0}%
\end{equation}%
\begin{equation}
\omega_{1}=\left(
\begin{array}
[c]{cccccc}%
0 & \tfrac{1}{2}p_{2} & \tfrac{1}{2}p_{3} & \tfrac{1}{2}q_{1} & \tfrac{1}{2}q_{2} & q_{3}\\
-\tfrac{1}{2}p_{2} & 0 & 0 & -1 & 0 & \tfrac{1}{2}q_{2}\\
-\tfrac{1}{2}p_{3} & 0 & 0 & 0 & -1 & \tfrac{1}{2}q_{1}\\
-\tfrac{1}{2}q_{1} & 1 & 0 & 0 & 0 & 0\\
-\tfrac{1}{2}q_{2} & 0 & 1 & 0 & 0 & 0\\
-q_{3} & -\tfrac{1}{2}q_{2} & -\tfrac{1}{2}q_{1} & 0 & 0 & 0
\end{array}
\right)  ,\label{p1}%
\end{equation}
where $I_{3}$ is an $3\times3$ unit matrix, the expansion coefficients
$F_{i}^{(1,1)}\equiv F_{i}$:
\[
F_{1}=-q_{1},\ \ F_{2}=-\tfrac{1}{4}q_{1}^{2}-q_{2}%
,\ \ \ F_{3}=-\tfrac{1}{2}q_{1}q_{2}-q_{3}%
\]
and Hamiltonian vector fields $x_{i}=\pi_{0}dh_{i},\ i=1,2,3$.

On the extended phase space $\mathcal{M}$ of dimension seven, with an
additional coordinate $c$, the differentials $dh_{i}$ form a
bi-inverse-Hamiltonian chain
\[%
\begin{array}
[c]{l}%
\Omega_{0}Y_{0}=0\\
\Omega_{0}Y_{1}=dh_{1}=\Omega_{1}Y_{0}\\
\Omega_{0}Y_{2}=dh_{2}=\Omega_{1}Y_{1}\\
\Omega_{0}Y_{3}=dh_{3}=\Omega_{1}Y_{2}\\
\qquad\quad\quad\,\,\,0=\Omega_{1}Y_{3},\,\,
\end{array}
\]
with presymplectic forms
\[
\Omega_{0}=\left(
\begin{array}
[c]{c|c}%
\omega_{0} & 0\\\hline
0 & 0
\end{array}
\right)  \quad,\quad\Omega_{1}=\left(
\begin{array}
[c]{c|c}%
\omega_{1} & dh_{1}\\\hline
-dh_{1}^{T} & 0
\end{array}
\right)
\]
d-compatible with respect to $\Pi_0$ and vector fields
\[
Y_{0}=(0,...,0,1)^{T},\ \ Y_{i}=X_{i}-F_{i}Y_{0},\ \ \ i=1,2,3,
\]
where $X_i=\Pi_0 dh_i$.
\vspace{0.3cm}\newline\textbf{Example 2.}\newline Consider the separation
relations on a six-dimensional phase space given by the following bare
separation curve
\[
\lambda^{2}(h_{1}^{(1)}\lambda+h_{2}^{(1)})+h_{1}^{(2)}=\tfrac{1}{2}\mu
^{2}\label{7.21c}%
\]
representing geodesic motion for a classical St\"{a}ckel system (this time of
non-Benenti type \cite{mac2005}). Using the coordinates, the Hamiltonians, and
the functions $\sigma_{i}$ from the previous example we find that
\begin{align*}
h_{1}^{(1)} &  =-\tfrac{1}{\sigma_{2}}h_{2},\\
h_{2}^{(1)} &  =h_{1}-\tfrac{\sigma_{1}}{\sigma_{2}}h_{2},\\
h_{1}^{(2)} &  =h_{3}-\tfrac{\sigma_{3}}{\sigma_{2}}h_{2}%
\end{align*}
and thus we see that the Hamiltonians $h_{i}^{(k)}$ are related to $h_{j}$
through the so-called generalized St\"{a}ckel transform (see \cite{bs} for
further details on the latter). They admit a quasi bi-inverse-Hamiltonian
representation (\ref{0.9})
\begin{align*}
\omega_{1}x_{1}^{(1)} &  =\omega_{0}x_{2}^{(1)}+F_{1}^{(1,1)}\omega_{0}%
x_{1}^{(1)}+F_{1}^{(1,2)}\omega_{0}x_{1}^{(2)},\ \\
\omega_{1}x_{2}^{(1)} &  =F_{2}^{(1,1)}\omega_{0}x_{1}^{(1)}+F_{2}%
^{(1,2)}\omega_{0}x_{1}^{(2)},\\
\omega_{1}x_{1}^{(2)} &  =F_{1}^{(2,1)}\omega_{0}x_{1}^{(1)}+F_{1}%
^{(2,2)}\omega_{0}x_{1}^{(2)}%
\end{align*}
with the presymplectic forms (\ref{p0}), (\ref{p1}), the expansion
coefficients
\[
F_{1}^{(1,1)}=-\sigma_{1}+\tfrac{\sigma_{3}}{\sigma_{2}},\ \ F_{2}%
^{(1,1)}=-\sigma_{2}+\tfrac{\sigma_{1}\sigma_{3}}{\sigma_{2}},\ \ \ F_{1}%
^{(2,1)}=\tfrac{\sigma_{3}^{2}}{\sigma_{2}},\
\]%
\[
F_{1}^{(1,2)}=-\tfrac{1}{\sigma_{2}},\ \ \ F_{2}^{(1,2)}=-\tfrac{\sigma_{1}%
}{\sigma_{2}},\ \ \ F_{1}^{(2,2)}=-\tfrac{\sigma_{3}}{\sigma_{2}}%
\]
and Hamiltonian vector fields $x_{i}^{(k)}=\pi_{0}dh_{i}^{(k)}$.

On the extended phase space $\mathcal{M}$ of dimension eight, with an
additional coordinates $c_{1}$ and $c_{2}$, the differentials $dh_{i}^{(k)}$
form a bi--inverse-Hamiltonian chains (\ref{0.13})
\[%
\begin{array}
[c]{ccc}%
\begin{array}
[c]{l}%
\Omega_{0}Y_{0}^{(1)}=0\\
\Omega_{0}Y_{1}^{(1)}=dh_{1}^{(1)}=\Omega_{1}Y_{0}^{(1)}\\
\Omega_{0}Y_{2}^{(1)}=dh_{2}^{(1)}=\Omega_{1}Y_{1}^{(1)}\\
\qquad\qquad\qquad0=\Omega_{1}Y_{2}^{(1)}%
\end{array}
&  &
\begin{array}
[c]{l}%
\Omega_{0}Y_{0}^{(2)}=0\\
\Omega_{0}Y_{1}^{(2)}=dh_{1}^{(2)}=\Omega_{1}Y_{0}^{(2)}\\
\qquad\qquad\qquad0=\Omega_{1}Y_{1}^{(2)},
\end{array}
\end{array}
\]
with the presymplectic forms
\[
\Omega_{0}=\left(
\begin{array}
[c]{c|c}%
\omega_{0} & 0\ \ 0\\\hline%
\begin{array}
[c]{c}%
0\\
0
\end{array}
& 0
\end{array}
\right)  \text{ \ , \ }\Omega_{1}=\left(
\begin{array}
[c]{c|c}%
\omega_{1} & dh_{1}^{(1)}\ \ \ dh_{1}^{(2)}\\\hline%
\begin{array}
[c]{c}%
-(dh_{1}^{(1)})^{T}\\
-(dh_{1}^{(2)})^{T}%
\end{array}
& 0
\end{array}
\right)
\]
d-compatible with respect to $\Pi_0$ and vector fields
\[
Y_{0}^{(1)}=(0,...,0,1,0)^{T},\ \ \ Y_{0}^{(2)}=(0,...,0,0,1)^{T}%
,\ \ Y_{1}^{(1)}=X_{1}^{(1)}-F_{1}^{(1,1)}Y_{0}^{(1)}-F_{1}^{(1,2)}Y_{0}%
^{(2)},\
\]%
\[
Y_{2}^{(1)}=X_{2}^{(1)}-F_{2}^{(1,1)}Y_{0}^{(1)}-F_{2}^{(1,2)}Y_{0}%
^{(2)},\ \ \ Y_{1}^{(2)}=X_{1}^{(2)}-F_{1}^{(2,1)}Y_{0}^{(1)}-F_{1}%
^{(2,2)}Y_{0}^{(2)}.
\]
\vspace{0.3cm}\newline\textbf{Example 3.}\newline Consider separation
relations on a six-dimensional phase space given by the following bare
separation curve, cubic in momenta,
\[
h_{1}^{(1)}\mu+h_{1}^{(2)}\lambda+h_{2}^{(2)}=\mu^{3}.
\]
The transformation $(\lambda,\mu)\rightarrow(q,p)$ to new canonical
coordinates in which all Hamiltonians are of a polynomial form is obtained
from the following two transformations:
\begin{eqnarray*}
u_1 &=& 3q_2-3q_3,\\
u_2 &=& -q_1p_2-q_1p_3+3q_3^2+5q_1^3-6q_2q_3,\\
u_3 &=& -q_3^3-9q_1^3q_3+q_1q_3p_2+q_1q_3p_3-6q_1^3q_2\\ &&+q_1^2p_1+3q_2q_3^2,\\
v_1 &=& -\frac{1}{q_1},\\
v_2 &=& \frac{3q_2-2q_3}{q_1},\\
v_3 &=&
p_3+\tfrac{2}{3}p_2-\frac{q_3^2}{q_1}+3\frac{q_2q_3}{q_1}-4q_1^2,
\end{eqnarray*}
and
\begin{eqnarray*}
u_1 &=& \lambda_1+\lambda_2+\lambda_3,\\
u_2 &=& \lambda_1\lambda_2+\lambda_1\lambda_3+\lambda_2\lambda_3,\\
u_3 &=& \lambda_1\lambda_2\lambda_3,\\
\mu_i &=& v_1\lambda_i^2+v_2\lambda_i+v_3, \qquad i=1,2,3.
\end{eqnarray*}

In the $(q,p)$-coordinates the Hamiltonians take the form
\begin{align*}
h_{1}^{(1)}  &  =p_{2}p_{3}+\tfrac{1}{3}p_{2}^{2}+p_{3}^{2}-7q_{1}^{2}%
p_{3}-4q_{1}^{2}p_{2}-3q_{2}p_{1}+18q_{1}q_{2}^{2}+13q_{1}^{4}+12q_{3}%
q_{1}q_{2},\\
h_{1}^{(2)}  &  =12q_{1}^{3}q_{2}+8q_{1}^{3}q_{3}-2q_{1}^{2}p_{1}%
+(-6q_{1}q_{2}-4q_{1}q_{3})p_{3}+p_{1}p_{3},\\
h_{2}^{(2)}  &  =\tfrac{1}{3}p_{2}p_{3}^{2}+\tfrac{1}{3}p_{2}^{2}p_{3}%
+\tfrac{2}{27}p_{2}^{3}-q_{1}^{2}p_{3}^{2}-\tfrac{4}{3}q_{1}^{2}p_{2}%
^{2}-q_{2}p_{1}p_{2}-q_{1}p_{1}^{2}-\tfrac{10}{3}q_{1}^{2}p_{3}p_{2}\\
&  +(q_{3}-3q_{2})p_{1}p_{3}+(21q_{1}^{2}q_{2}+6q_{3}q_{1}^{2})p_{1}%
+(4q_{3}q_{1}q_{2}+6q_{1}q_{2}^{2}+\tfrac{22}{3}q_{1}^{4})p_{2}\\
&  +(7q_{1}^{4}+18q_{1}q_{2}^{2}+6q_{3}q_{1}q_{2}-4q_{1}q_{3}^{2})p_{3}%
-8q_{1}^{3}q_{3}^{2}-72q_{3}q_{1}^{3}q_{2}-90q_{1}^{3}q_{2}^{2}-12q_{1}^{6}.
\end{align*}
They form a quasi bi-inverse-Hamiltonian chain (\ref{0.9})
\begin{align*}
\omega_{1}x_{1}^{(1)}  &  =F_{1}^{(1,1)}\omega_{0}x_{1}^{(1)}+F_{1}%
^{(1,2)}\omega_{0}x_{1}^{(2)},\\
\omega_{1}x_{1}^{(2)}  &  =\omega_{0}x_{2}^{(2)}+F_{1}^{(2,1)}\omega_{0}%
x_{1}^{(1)}+F_{1}^{(2,2)}\omega_{0}x_{1}^{(2)},\\
\omega_{1}x_{2}^{(2)}  &  =F_{2}^{(2,1)}\omega_{0}x_{1}^{(1)}+F_{2}%
^{(2,2)}\omega_{0}x_{1}^{(2)},
\end{align*}
with the non-canonical symplectic form%
\[
\omega_{1}=\left(
\begin{array}
[c]{cccccc}%
0 & -B & -C & q_{3} & -A & -2q_{1}^{2}\\
B & 0 & 24q_{1}^{2} & -3q_{1} & -3q_{2}+q_{3} & 0\\
C & -24q_{1}^{2} & 0 & 2q_{1} & q_{2} & q_{3}\\
-q_{3} & 3q_{1} & -2q_{1} & 0 & 0 & 0\\
A & 3q_{2}-q_{3} & -q_{2} & 0 & 0 & \tfrac{1}{3}q_{1}\\
2q_{1}^{2} & 0 & -q_{3} & 0 & -\tfrac{1}{3}q_{1} & 0
\end{array}
\right)  ,
\]
where $A=\tfrac{1}{3}p_{2}+\tfrac{1}{3}p_{3}-3q_{1}^{2}$, $B=54q_{1}%
q_{2}+24q_{1}q_{3}-3p_{1}$, $C=-24q_{1}q_{2}-12q_{1}q_{3}+p_{1}$ and the
expansion coefficients
\[
F_{1}^{(1,1)}=-q_{3},\ \ F_{1}^{(1,2)}=-q_{1},\ \ \ F_{1}^{(2,1)}=-\tfrac
{1}{3}p_{2}+q_{1}^{2},\ \ \ F_{1}^{(2,2)}=-2q_{3}+3q_{2},
\]%
\[
F_{2}^{(2,1)}=5q_{3}q_{1}^{2}+6q_{1}^{2}q_{2}-q_{1}p_{1}-\tfrac{1}{3}%
q_{3}p_{2},\ \ \ F_{2}^{(2,2)}=-4q_{1}^{3}-q_{3}^{2}+3q_{2}q_{3}+\tfrac{2}%
{3}q_{1}p_{2}+q_{1}p_{3}.
\]

On the extended phase space $\mathcal{M}$ of dimension eight, with additional
coordinates $c_{1}$ and $c_{2}$, the differentials $dh_{i}^{(k)}$ form a
bi--inverse-Hamiltonian chains (\ref{0.13})
\[%
\begin{array}
[c]{ccc}%
\begin{array}
[c]{l}%
\Omega_{0}Y_{0}^{(1)}=0\\
\Omega_{0}Y_{1}^{(1)}=dh_{1}^{(1)}=\Omega_{1}Y_{0}^{(1)}\\
\qquad\qquad\qquad0=\Omega_{1}Y_{1}^{(1)},
\end{array}
&  &
\begin{array}
[c]{l}%
\Omega_{0}Y_{0}^{(2)}=0\\
\Omega_{0}Y_{1}^{(2)}=dh_{1}^{(2)}=\Omega_{1}Y_{0}^{(2)}\\
\Omega_{0}Y_{2}^{(2)}=dh_{2}^{(2)}=\Omega_{1}Y_{1}^{(2)}\\
\qquad\qquad\qquad0=\Omega_{1}Y_{2}^{(2)}%
\end{array}
\end{array}
\]
with the presymplectic forms
\[
\Omega_{0}=\left(
\begin{array}
[c]{c|c}%
\omega_{0} & 0\ \ 0\\\hline%
\begin{array}
[c]{c}%
0\\
0
\end{array}
& 0
\end{array}
\right)  \text{ \ , \ }\Omega_{1}=\left(
\begin{array}
[c]{c|c}%
\omega_{1} & dh_{1}^{(1)}\ \ \ dh_{1}^{(2)}\\\hline%
\begin{array}
[c]{c}%
-(dh_{1}^{(1)})^{T}\\
-(dh_{1}^{(2)})^{T}%
\end{array}
& 0
\end{array}
\right)
\]
d-compatible with respect to $\Pi_0$ and vector fields
\[
Y_{0}^{(1)}=(0,...,0,1,0)^{T},\ \ \ Y_{0}^{(2)}=(0,...,0,0,1)^{T}%
,\ \ \ Y_{1}^{(1)}=X_{1}^{(1)}-F_{1}^{(1,1)}Y_{0}^{(1)}-F_{1}^{(1,2)}%
Y_{0}^{(2)},\
\]%
\[
Y_{1}^{(2)}=X_{1}^{(2)}-F_{1}^{(2,1)}Y_{0}^{(1)}-F_{1}^{(2,2)}Y_{0}%
^{(2)},\ \ \ \ Y_{2}^{(2)}=X_{2}^{(2)}-F_{2}^{(2,1)}Y_{0}^{(1)}-F_{2}%
^{(2,2)}Y_{0}^{(2)}.
\]

The bi-Hamiltonian extensions of systems from presented examples the reader
can find in \cite{bb1}.

\section*{Acknowledgement}

The work is partially supported by Polish MNiSW research grant no. N N202 404933

\end{document}